\documentstyle[12pt]{article}

\font\sm=cmr9
\newcommand {\be}{\begin{equation}} 
\newcommand{\fe}{\end{equation}}
\newcommand{\eqn}{\label}
\def\dalemb#1#2{{\vbox{\hrule height .#2pt \hbox{\vrule width.#2pt height#1pt \kern#1pt\vrule width.#2pt}\hrule height.#2pt}}}
\def\square{\mathord{\dalemb{5.9}{6}\hbox{\hskip1pt}}}

\begin{document} 
  
\title{\bf NON-DIVERGENCE OF GRAVITATIONAL SELF INTERACTIONS FOR NAMBU-GOTO
STRINGS} 

\author{Brandon Carter{$^{1}$} and Richard A. Battye{$^{2}$}\\ \\ \it
${}^{1}$ Department d'Astrophysique Relativiste et de Cosmologie,\\ \it Centre National
de la Recherche Scientific, \\ \it Observatoire de Paris, 92195 Meudon
Cedex, France. \\ \\ \it ${}^2$ Department of Applied Mathematics and
Theoretical Physics, \\ \it University of Cambridge, \\ \it Silver Street,
Cambridge CB3 9EW, U.K. } 

\maketitle

\begin{abstract}   

The classical linearised gravitational self interaction of a
Nambu-Goto string is examined in four spacetime dimensions. Using a
conveniently gauge independent tensorial treatment, the divergent part
of the self-force  is shown to be exactly zero. This is due to
cancelation by a contribution that was neglected in the previous
treatments.  This result has implications for many applications.

\end{abstract}

Relativistic string models are of importance in many high energy
physics contexts.  Cosmic strings formed at a cosmological phase
transitions may have been the primordial seeds for galaxy formation~\cite{VS94},
while fundamental strings at the Planck scale are at the heart of
Superstring theory~\cite{GSW} and the duality driven revolution known as
M-theory.  The simplest prototype in all these contexts is the Nambu-Goto
model.  However, even at  classical level the fundamental self
interaction properties of Nambu-Goto strings are still not properly
understood.

Long before Dirac's definitive study~\cite{dirac} in the 1930's it was already well
known that the electromagnetic self interactions of a classical point
particle gave rise to a divergent self energy,
 but that this could be renormalized into the mass, leaving a finite
radiation reaction effect. The same basic concepts have since been
found to be applicable to the self interactions of particles, and of strings, 
when coupled to a variety of fields including gravity~\cite{QS90,CHH90}.However 
the evaluation of gravitational self interactions is rather
delicate, one of reasons being that the metric is often hidden, for
example, in the induced metric of the worldsheet. In this work, we use
powerful gauge independent tensorial notation, in which the dependence
on the metric is explicit, to re-compute the divergent part of
gravitational self interaction for linearized gravity. Contrary to what emerged
from previous treatments of this problem, we find that it is zero for
Nambu-Goto strings, though it would be non-zero for strings models~\cite{C95}
of a more general kind.

Strings can be described by spacetime coordinates
$x^{\mu}=X^{\mu}(\sigma^a)$, where $a=0,1$ and $\sigma^a$ are the
internal coordinates on the 2-dimensional worldsheet. The induced
metric on this worldsheet is then \be
\gamma_{ab}=g_{\mu\nu}\partial_aX^{\mu}\partial_bX^{\nu}\,,\eqn{1} \fe
and its energy-momentum tensor is \be \hat T{^{\mu\nu}}(x^{\mu})=\Vert
g\Vert^{-1/2}\int \overline T{^{\mu\nu}}\, \delta^{\rm
4}[x-X(\sigma^a)]\, \Vert\gamma \Vert^{1/2}\, d^{2}\sigma\,, \eqn{2}
\fe where $\Vert g\Vert$ is the determinant of the spacetime metric,
$\Vert\gamma\Vert$ is that of the induced metric and  ${\overline
T}{^{\mu\nu}}$ is the regular worldsheet supported tensor field
representing the surface energy-momentum density.

One can define the fundamental tensor of the worldsheet, that is, the
spacetime projection of the internal metric tensor $\gamma^{ab}$, which
is given by \be
\eta^{\mu\nu}=\gamma^{ab}\partial_aX^{\mu}\partial_bX^{\nu}\,.\eqn{7}
\fe 
This fundamental tensor allows one to set up a convenient formalism for
describing the dynamics strings in a manner that is independent
the choice of the internal coordinate gauge. 
Firstly, one can define its orthogonal
complement 
\be \perp_{\mu\nu}=g_{\mu\nu}-\eta_{\mu\nu}\fe
 and the tangentially projected differentiation operator
${\overline\nabla}_{\mu}=\eta_{\mu}^{\ \nu}\nabla_{\nu}$ for tensor
fields whose support is confined to the worldsheet. Using this
projected differentiation one can define the second fundamental tensor
and curvature vector by \be
K_{\mu\nu}^{\ \ \rho}=\eta_{\sigma\nu}{\overline
\nabla}_{\mu}\eta^{\sigma\rho}\,,\quad
K^{\rho}=g^{\mu\nu}K_{\mu\nu}^{\ \ \rho}\,.  \fe Using the
orthogonality of $\eta_{\mu\nu}$ and $\perp_{\mu\nu}$, one can derive
the following elementary properties of the second fundamental tensor:
\begin{eqnarray} & \perp^{\sigma\mu}K_{\mu\nu}^{\ \ \rho}=0\,,\quad
\eta_{\sigma\rho}K_{\mu\nu}^{\ \ \rho}=0\,,\quad
K_{[\mu\nu]}^{\ \ \ \ \rho}=0\,, & \\ &
{\overline\nabla}_{\mu}\eta_{\nu\rho}=2K_{\mu(\nu\rho)}\,,\quad
{\overline\nabla}_{\mu}\eta^{\mu\rho}=K^{\rho}\,,& \end{eqnarray} where
round and square brackets denote index symmetrization and
antisymmetrization repsectively.

The simplest kind of model is governed by the Nambu-Goto action, which
takes the simple form \be {\cal I}_{_{\rm GN}}=-m^2 \int\,\Vert\gamma\Vert^{1/2}\, d^2\sigma \,,\eqn{4} \fe where $m$ is a
constant, having the dimensions of mass. In the case of a cosmic string
model providing a macroscopic description of a vortex defect of the
vacuum, $m$ would typically be of the order of magnitude of the Higgs
mass scale associated with relevant spontaneus symmetry breaking
mechanism and $m^2$ would correspond to the mass per unit length. For
this model the relevant surface energy-momentum tensor that is required
for substitution in (\ref{2}) will be given simply by \be \overline
T{^{\mu\nu}}=-T\,\eta^{\mu\nu}\,,\eqn{5} \fe where the scalar $T$ is
the string tension, which will be given by 
\be T=m^2\, .\fe

Using the tensorial notation described above the equation of motion for
the free Nambu-Goto string is just \be TK^{\rho}=0\,, \fe 
which is the analogue of Newton's 2nd law of motion.
In the more complicated worldsheet notation 
\be
K^{\rho}=\Vert\gamma\Vert^{-1/2}\partial_a\left(\Vert\gamma\Vert^{1/2}
\gamma^{ab} \partial_bX^{\rho}\right)+\Gamma^{\rho}_{\ \alpha\beta}
\gamma^{ab}\partial_aX^{\alpha}\partial_bX^{\beta}\, .
\fe
In Minkowski coordinates  using the conformal gauge, with $\dot{}=
\partial/\partial\tau$,
${}^{\prime}=\partial/\partial\sigma$ where $\tau=\sigma^{0}$
and $\sigma=\sigma^{1}$ is the space coordinate along the string, the
curvature vector has the form $K^\mu=\Vert\gamma\Vert^{-1/2}(\ddot
X^{\mu}-X^{\mu\prime\prime})$ so the equation of motion reduces to the well
 known form $ \ddot
X^{\mu}-X^{\mu\prime\prime}=0\, .$ 

The linearized Einstein equations describe the coupling of the string
to gravity in the weak field limit. Using a perturbation
$\delta g_{\mu\nu}=h_{\mu\nu}$ and the usual De Donder gauge condition
$\nabla^\mu h_{\mu\nu}={1\over 2}\nabla_\nu h$ with
$h=g^{\mu\nu}h_{\mu\nu}$, the Dalembertian wave equation is \be \square
h^{\mu\nu}=-8\pi  \hbox{\sm G} \big(2\hat T^{\mu\nu} -\hat T
g^{\mu\nu}\big)\,, \eqn{3} \fe where
$\square=g^{\mu\nu}\nabla_{\!\mu\!}\nabla_{\!\nu}$, $\hat
T=g^{\mu\nu}\hat T_{\mu\nu}$ and {\sm G} is Newton's constant.  As
usual, when dealing with a singular source distribution, for which the
self interaction contribution will be divergent on the worldsheet, it
will be convenient to consider the total field as the sum of a
divergent local contribution, $\widehat h_{\mu\nu}$ say, and a finite
remainder $\widetilde h_{\mu\nu}$ due to long range interactions and
passing radiation from outside the system, in the form \be
h_{\mu\nu}=\widehat h_{\mu\nu}+\widetilde h_{\mu\nu}\,.\eqn{8} \fe In
four spacetime dimensions, the divergence of a point particle would be
a pole, while in the case of a string it will be logarithmic.
Therefore, if we introduce a short range cut-off $\delta_\ast$ and also
one at long distances $\Delta$, then divergence of the point particle
is dominated by the short range, ultra-violet cut-off, while the
logarithmic nature of the divergence in the case of a string, makes the
strength of divergence dependent on the ratio of the two cut-offs.
In the case of a string, the short range cut off $\delta_\ast$ is
provided by the width of the string, which for a cosmic string will
typically be of order $m^{-1}$ assuming natural units, while the long
range cut-off will be provided by the radius of curvature or
inter-string separation. In higher dimensions the situation is more
complicated, but nonetheless it should be possible to take into account
the relevent divergences by some function of $\delta_\ast$ and
$\Delta$.

The standard Green function solution of (\ref{3}) provides an estimate
for the divergent part of the solution which is of the form \be
\widehat h_{\mu\nu}=-2\hbox{\sm G}\,\big(2\overline T_{\!\mu\nu}-
{\overline
T}g_{\mu\nu}\big)\log\left(\Delta^2/\delta_*^2\right)\,,\eqn{9} \fe
where ${\overline T}=g^{\mu\nu}{\overline T}^{\mu\nu}$. Using the
specific expression for a Nambu-Goto string this becomes 
\be \widehat h_{\mu\nu}=-4\hbox{\sm G}\, T\perp_{\mu\nu}\log\left(
\Delta^2/\delta_\ast^2\right)\, , \eqn{10}\fe 
whose important feature is that
it is proportional to the orthogonal projection operator $\perp_{\mu\nu}$.
This orthogonality property is the special feature of Goto Nambu strings
(as opposed to point particles, membranes, or more general string models)
which is ultimately responsible for the cancellation to be described below.

To evaluate the dynamical effect of the perturbing field it is
also necessary to know its gradient. By a standard calculation, of which
simpler illustrations are provided by the cases of scalar~\cite{K93}
and electromagnetic~\cite{C97} interactions, it can be seen that for 
the divergent part given by (\ref{9}), the corresponding regularised
limit of the gradient is expressible as 
\be
\widehat{\nabla_{\!\rho} h_{\mu\nu}} =\overline\nabla_{\!\rho} \widehat
h_{\mu\nu} +{_1\over^2} K_\rho\widehat h_{\mu\nu}\,.\eqn{12}
\fe  

Up to this point everything we have written here has long been  well 
known and generally agreed~\cite{VS94}, the only innovation being its 
translation into more modern and streamlined tensorial notation. 
Our difference from preceeding analysis arises when we come to the
relevant gravitationally perturbed dynamical equations. 

To get the corresponding gravitationally perturbed dynamical equations, we 
start from the appropriate gravitationally perturbed action
\be {\cal I} = -T\int\, \big(1+{_1\over^2}\eta^{\mu\nu}h_{\mu\nu}\big)
 \Vert\gamma\Vert^{1/2}\, d^2\sigma \, ,\eqn{17}\fe
that is obtained by making the linearized substitution 
$g_{\mu\nu}\mapsto g_{\mu\nu}+h_{\mu\nu}$ in (\ref{4}). 
Requiring (\ref{17}) to be locally invariant with respect to small 
variations of the world sheet leads  to dynamical equation of the 
standard form
\be T K^\rho = - f_{\rm g}{^\rho}\, ,\eqn{18}\fe
where the  $f_{\rm g}{^\rho}$ is the
effective gravitational force density vector acting on the wordsheet.

It is in the evaluation of this force density that we deviate from
the traditional formula (see, for example, ref.~\cite{VS94}), which is 
expressible simply as
\be  f_{_{\rm trad}}^{\, \rho}=
T \,\eta^{\mu\nu}\big{(} \nabla_{\mu}h_\nu^{\ \rho}-
\textstyle{1\over 2}\nabla^{\rho}h_{\mu\nu}\big{)}\, .\eqn{22}\fe
The use of this formula may be justified if it is understood to describe
not a force in the strict vectorial sense but a pseudo force applicable
in a special gauge, very similar to, but
not exactly the conformal gauge. However, such an approach involves delicate subtleties that can be and have been a source of error. Using a more robust strictly tensorial treatment that does not depend on any particular internal gauge (conformal or otherwise) in the string, we actually obtain~\cite{BC95}, by proceeding carefully step by step through the variation of (\ref{17}), an
expression of the form
\be
f_{\rm g}{^\rho}= f_{_{\rm I}}{^\rho}+ f_{_{\rm I\! I}}{^\rho}\,,\eqn{19}
\fe
with
\be
f_{_{\rm I}}{^\rho}=\perp^{\!\rho}_{\ \nu} f_{_{\rm trad}}^{\, \nu}
\, ,\eqn{20}
\fe
and 
\be 
f_{_{\rm I\! I}}{^\rho}= T\,
\big{(}\! \perp^{\rho\nu}K^{\mu}+\textstyle{1\over
2}\eta^{\mu\nu}K^{\rho}-K^{\mu\nu\rho}\big{)}h_{\mu\nu}\,.\eqn{21}
\fe

The use -- in place of the true force $f_{\rm g}{^\rho}$ in (\ref{18}) -- of the  traditional pseudo force $f_{_{\rm trad}}^{\, \rho}$ given by (\ref{22}) as if it were a genuinely vectorial quantity would be not just physically 
inexact but also mathematically inadmissible, since it would lead to an overdetermined system of equations of motion. The problem is analogous to the that which arises when considering the effect of a scalar field $\phi$ on a point particle with unit 4-velocity
$u^\mu$ and acceleration $a^\mu=u^\nu\nabla_{\!\nu} u^\mu$. If, instead of the
mathematically admissible equation of motion $a^\mu=\nabla^\mu\phi+ u^\mu
u_\nu\nabla^\nu\phi$, one simply took $a^\mu=\nabla^\mu\phi$,  it would be
incompatible with the identity $a_\mu u^\mu=0$.

In order to be consistent 
with the orthogonality property 
\be K^\rho \eta_\rho^{\ \sigma}=0 \, ,\eqn{23}\fe
of the curvature vector, which 
follows~\cite{BC95}, as a mathematical identity,
it is necessary that the force term on the right of (\ref{18}), should 
have a similar orthogonality property. This requirement of vanishing 
tangential projection is automatically  satisfied by the separate 
contributions $f_{_{\rm I}}{^\rho}$ and $f_{_{\rm I\! I}}{^\rho}$, 
and hence by $f_{\rm g}{^\rho}$, but it is not satisfied 
by $f_{_{\rm trad}}^{\, \rho}$. 

The reason why the problem with (\ref{22}) was not noticed
earlier may be that the special, purely orthogonal, form of (\ref{10})
ensures that the divergent contribution $\widehat h_{\mu\nu}$ does not
engender any unacceptable tangential component in the corresponding
divergent force term, which is obtainable (using (\ref{12}) or
otherwise) in the well known form
\be \widehat f{_{_{\rm
trad}}^{\, \rho}} = -4 \hbox{\sm G}\, T^2\, K^\rho\, \log\left(
\Delta^2 /\delta_\ast^{\, 2}\right)\,.\eqn{24}  \fe 
Unlike the corresponding finite part 
$\widetilde f{_{_{\rm trad}}^{\, \rho}}$, this does after 
all satisfy the orthogonality requirement. This means 
that for the divergent contribution from $f_{_{\rm I}}{^\rho}\,$ 
the orthogonal projection in (\ref{20}) will have no effect, so it
will be given by the same formula, 
\be \widehat f{_{_{\rm I}}{^\rho}}=
 \widehat f{_{_{\rm trad}}^{\, \rho}}\, .\eqn{25}\fe 
The formula (\ref{24}) is what is traditionally
quoted~\cite{QS90,CHH90,VS94} for the divergent part of the
 gravitational self force and it is evidently interpretable as
representing the effect of a tension adjustment $\Delta_{_{\rm I}}T =-4
\hbox{\sm G}\, T^2\, \log\left( \Delta^2 /\delta_\ast^{\, 2}\right)$.

We now come to the crucial point, which is that this
adjustment will be precisely cancelled by the corresponding
contribution $\Delta_{_{\rm I\! I}}T$ from the previously overlooked
force term $f_{_{\rm I\! I}}{^\rho}$, whose divergent part can be seen,
by substituting (\ref{10}) in (\ref{21}), to be given by
\be \widehat f{_{_{\rm I\! I}}{^\rho}} = 
-\widehat f{_{_{\rm I}}{^\rho}}\, .\eqn{26}\fe
Thus the total force density due to the divergent part of the 
self field  will vanish,  
\be \widehat f{_{\rm g}{^\rho}} =0\, ,\eqn{27}\fe 
so  no net renormalisation is needed after all:
$\Delta_{\rm g} T= \Delta_{_{\rm I}}T+\Delta_{_{\rm I\! I}}T =0$. 
The practical implication is that the dynamical equations
will be obtainable {\it in a well behaved form }  just by replacing 
 $f{_{\rm g}{^\rho}}$ in (\ref{18}) by its finite part
$\widetilde f{_{\rm g}{^\rho}}$, which is to be
 evaluated by retaining just the finite part of $h_{\mu\nu}$ in (\ref{20}) and (\ref{21}), that is, the contribution from $\widetilde h_{\mu\nu}$.

It has been argued~\cite{DH89} in the context of quantised superstring 
theory that there should be no string tension renormalisation in the
relevant low energy classical limit, which involves not just gravitational
coupling but also dilatonic and axonic contributions. If this is
correct, then our theorem to the effect that the gravitational part vanishes
by itself implies, as a corollary, that the appropriately calibrated 
axionic and dilatonic parts should be mutually cancelling, a prediction that
has recently been confirmed directly~\cite{BD98}.

Although absent in the Nambu-Goto case, a non trival renormalisation
will be needed for the transonic string
model~\cite{C95} that represents the macroscopically averaged effect
of wiggles in an underlying Goto-Nambu model. A preliminary analysis of the 
general case has already been undertaken~\cite{C98} and a more
complete treatment is in preparation.

The authors wish to thank Thibault Damour, Patrick Peter, and Paul Shellard
for instructive conversations. RAB is funded by Trinity College.

\def\jnl#1#2#3#4#5#6{\hang{#1, {\it #4\/} {\bf #5}, #6 (#2).}}
\def\jnltwo#1#2#3#4#5#6#7#8{\hang{#1, {\it #4\/} {\bf #5}, #6; {\it
ibid} {\bf #7} #8 (#2).}} 
\def\prep#1#2#3#4{\hang{#1, #4.}} 
\def\proc#1#2#3#4#5#6{{#1 [#2], in {\it #4\/}, #5, eds.\ (#6).}}
\def\book#1#2#3#4{\hang{#1, {\it #3\/} (#4, #2).}}
\def\jnlerr#1#2#3#4#5#6#7#8{\hang{#1 [#2], {\it #4\/} {\bf #5}, #6.
{Erratum:} {\it #4\/} {\bf #7}, #8.}}
\def\prl{Phys.\ Rev.\ Lett.}
\def\pr{Phys.\ Rev.}
\def\pl{Phys.\ Lett.}
\def\np{Nucl.\ Phys.}
\def\prp{Phys.\ Rep.}
\def\rmp{Rev.\ Mod.\ Phys.}
\def\cmp{Comm.\ Math.\ Phys.}
\def\mpl{Mod.\ Phys.\ Lett.}
\def\apj{Ap.\ J.}
\def\apjl{Ap.\ J.\ Lett.}
\def\aap{Astron.\ Ap.}
\def\cqg{Class.\ Quant.\ Grav.} 
\def\grg{Gen.\ Rel.\ Grav.}
\def\mn{MNRAS}
\def\ptp{Prog.\ Theor.\ Phys.}
\def\jetp{Sov.\ Phys.\ JETP}
\def\jetpl{JETP Lett.}
\def\jmp{J.\ Math.\ Phys.}
\def\zpc{Z.\ Phys.\ C}
\def\cupress{Cambridge University Press}
\def\pup{Princeton University Press}
\def\wss{World Scientific, Singapore}
\def\oup{Oxford University Press}

\end{document}